\RequirePackage{ifpdf}
\ifpdf 
\documentclass[pdftex]{sigma}
\else
\documentclass{sigma}
\fi

\usepackage{bm}

\begin{document}

\newcommand{\lplus}{\raisebox{-1.3mm}{\Huge $+$}}
\newcommand{\uvec}[1]{{\scriptsize $(#1)$}}
\newcommand{\emp}{\hspace{2.35mm}{\scriptsize$(0,0)$}\hspace{2.35mm}}
\newcommand{\empp}{\hspace{1.15mm}{\scriptsize$(0,0)$}\hspace{1.15mm}}

\allowdisplaybreaks

\renewcommand{\thefootnote}{$\star$}

\renewcommand{\PaperNumber}{028}

\FirstPageHeading

\ShortArticleName{Yang--Baxter Maps from the Discrete BKP Equation}

\ArticleName{Yang--Baxter Maps from the Discrete BKP Equation\footnote{This paper is a
contribution to the Proceedings of the Workshop ``Geometric Aspects of Discrete and Ultra-Discrete Integrable Systems'' (March 30 -- April 3, 2009, University of Glasgow, UK). The full collection is
available at
\href{http://www.emis.de/journals/SIGMA/GADUDIS2009.html}{http://www.emis.de/journals/SIGMA/GADUDIS2009.html}}}

\Author{Saburo KAKEI~$^\dag$, Jonathan J.C. NIMMO~$^\ddag$ and Ralph WILLOX~$^\S$}

\AuthorNameForHeading{S. Kakei, J.J.C. Nimmo and R. Willox}

\Address{$^\dag$~Department of Mathematics, College of Science, Rikkyo University,\\
\hphantom{$^\dag$}~3-34-1 Nishi-Ikebukuro, Toshima-ku, Tokyo 171-8501, Japan}
\EmailD{\href{mailto:kakei@rikkyo.ac.jp}{kakei@rikkyo.ac.jp}}
\URLaddressD{\url{http://www.rkmath.rikkyo.ac.jp/~kakei/index_e.html}}

\Address{$^\ddag$~Department of Mathematics, University of Glasgow, Glasgow G12 8QQ, UK}
\EmailD{\href{mailto:j.nimmo@maths.gla.ac.uk}{j.nimmo@maths.gla.ac.uk}}
\URLaddressD{\url{http://www.maths.gla.ac.uk/~jjcn/}}

\Address{$^\S$~Graduate School of Mathematical Sciences, The University of Tokyo,\\
\hphantom{$^\S$}~3-8-1 Komaba, Meguro-ku, Tokyo 153-8914, Japan}
\EmailD{\href{mailto:willox@ms.u-tokyo.ac.jp}{willox@ms.u-tokyo.ac.jp}}

\ArticleDates{Received November 13, 2009, in f\/inal form March 19, 2010;  Published online March 31, 2010}

\Abstract{We construct
rational and piecewise-linear Yang--Baxter maps for a general
$N$-reduction of the discrete BKP equation.}


\Classification{37K10; 37K60; 39A10; 39A14}

\section{Introduction}

Let $R$ be a map $R:\mathbb{C}^2\to\mathbb{C}^2$ and let
$R_{ij}:\mathbb{C}^n\to\mathbb{C}^n$ be the map that acts
as $R$ on the $i$-th and $j$-th factors in the direct product $\mathbb{C} \times \cdots \times \mathbb{C}$ and identically on the others.
A map $R$ is called a Yang--Baxter map if it satisf\/ies the
Yang--Baxter equation (with parameters) \cite{Veselov1,Veselov2},
\begin{gather}
\label{YBE}
R_{12}(\lambda_1,\lambda_2)R_{13}(\lambda_1,\lambda_3)
R_{23}(\lambda_2,\lambda_3)=R_{23}(\lambda_2,\lambda_3)
R_{13}(\lambda_1,\lambda_3)R_{12}(\lambda_1,\lambda_2).
\end{gather}

An important example of a Yang--Baxter map can be derived from the 3D-consistency condition for the discrete potential KdV equation \cite{PTV}:
\begin{gather}
R(\kappa,\mu): \
(u,v) \mapsto \left(vp, \frac{u}{p}\right),\qquad
p=\frac{1+\kappa uv}{1+\mu uv},
\label{dpKdV_map}
\end{gather}
which is the $H_{\rm III}^B$ map in the classif\/ication of quadrirational Yang--Baxter maps found in \cite{PSTV} (which is in fact equivalent to the $F_\mathrm{III}$ map in the Adler--Bobenko--Suris classif\/ication \cite{ABS}). An important property of the map \eqref{dpKdV_map} is that it is related to
a geometric crystal of $A_1^{(1)}$-type \cite{Etingof}.
Indeed, if we set
\begin{gather*}
u=e^{U/\epsilon},\qquad v=e^{V/\epsilon},\qquad p=e^{P/\epsilon},
\qquad \kappa=e^{-K/\epsilon},\qquad \mu=e^{-M/\epsilon},
\end{gather*}
and take the ultra-discrete limit (henceforth ``ud-limit'') $\epsilon\to+0$ \cite{TTMS},
we obtain a piecewise-linear Yang--Baxter map,
\begin{gather}
(U,V)\mapsto \left(V+P,U-P\right), \qquad
P=\max(0,U+V-K)-\max(0,U+V-M),
\label{A_1^{(1)}BBS}
\end{gather}
where we have used the formula
\begin{gather}
\lim_{\epsilon\to+0}
\epsilon\log\big(e^{A/\epsilon}+e^{B/\epsilon}\big)
=\max(A,B), \qquad A,B\in\mathbb{R}.
\end{gather}
It can be verif\/ied that \eqref{A_1^{(1)}BBS} coincides with
the combinatorial $R$-matrix of $A_1^{(1)}$-type \cite{HHIKTT},
and is in fact equivalent to the evolution rule of a
soliton cellular automata
(a box-ball system ``with a~carrier'')~\cite{TM}.
Furthermore, if we choose $\kappa=0$, the map~\eqref{dpKdV_map}
coincides with Hirota's discrete KdV equation~\cite{Hirota_dKdV}
and is therefore related to the Takahashi--Satsuma soliton cellular automata~\cite{TS} as discussed in~\cite{KNW}.

In a previous paper \cite{KNW}, we showed how several Yang--Baxter maps
can be directly obtained from the discrete KP hierarchy.
The aim of the present paper is to investigate the BKP counterpart of this construction and
of the map~\eqref{dpKdV_map} in particular.

\section[Hirota-Miwa equation and Harrison-type map]{Hirota--Miwa equation and Harrison-type map}
\label{section:dAKP}

In this section, we briefly describe a non-autonomous version of the method that enabled us to obtain
the map \eqref{dpKdV_map}
from the Hirota--Miwa equation,
\begin{gather}
(b-c)\tau(\ell+1,m,n)\tau(\ell,m+1,n+1)
+(c-a)\tau(\ell,m+1,n)\tau(\ell+1,m,n+1) \nonumber
\\
\qquad{}+(a-b)\tau(\ell,m,n+1)\tau(\ell+1,m+1,n)=0,
\label{HirotaMiwaEq}
\end{gather}
which was introduced by Hirota \cite{Hirota_DAGTE}.
The connection between \eqref{HirotaMiwaEq} and
the the $A$-type KP hierar\-chy was clarif\/ied by Miwa~\cite{Miwa}.
A non-autonomous version of the Hirota--Miwa equation,
\begin{gather}
\left\{b(m)-c(n)\right\}\tau(100)\tau(011)
+\left\{c(n)-a(\ell)\right\}\tau(010)\tau(101)
\nonumber\\
\qquad{}+\left\{a(\ell)-b(m)\right\}\tau(001)\tau(110)=0,
\label{na_HirotaMiwaEq}
\end{gather}
was studied in \cite{WTS1,WTS2}.
Here we have used the abbreviations,
$\tau(100)=\tau(\ell+1,m,n)$, $\tau(101)=\tau(\ell+1,m,n+1)$, etc.
The non-autonomous Hirota--Miwa (naHM) equation~\eqref{na_HirotaMiwaEq}
arises as the compatibility condition for the linear system
(cf.~\cite{DJM_II,Nimmo_dAKP} for autonomous case),
\begin{gather}
\psi(010)
=\frac{a(\ell)}{b(m)}\psi(100)
+\left\{1-\frac{a(\ell)}{b(m)}\right\}u(000)\psi(000),
\nonumber\\
\psi(001) =\frac{a(\ell)}{c(n)}\psi(100)
+\left\{1-\frac{a(\ell)}{c(n)}\right\}v(000)\psi(000),\label{linear_naHM}
\end{gather}
where $u(\ell,m,n)$, $v(\ell,m,n)$ are def\/ined as
\begin{gather}
\label{def:uv}
u(000)=\frac{\tau(000)\tau(110)}{\tau(100)\tau(010)},\qquad
v(000)=\frac{\tau(000)\tau(101)}{\tau(100)\tau(001)}.
\end{gather}

We now impose the following 2-reduction condition with respect to
the variable $\ell$;
\begin{gather}
u(\ell+2,m,n)=u(\ell,m,n),\qquad v(\ell+2,m,n)=v(\ell,m,n),
\nonumber\\
\psi(\ell+2,m,n)=\lambda\psi(\ell,m,n),\qquad
a(\ell+2)=a(\ell)\qquad \forall\,\ell\in\mathbb{Z}.
\label{2reduction_1}
\end{gather}
We furthermore require that the evolution in the remaining lattice directions is autonomous:
\begin{gather}
b(m)=b,\qquad c(n)=c \qquad \forall\, m,n\in\mathbb{Z}.
\label{2reduction_2}
\end{gather}
Then the linear system \eqref{linear_naHM} is
reduced to the $2\times 2$ system
\begin{gather}
\label{2x2LinearEqs_naHM_2-reduced}
\begin{bmatrix}
\psi_0(m+1,n)\\ \psi_1(m+1,n)
\end{bmatrix}
=\mathcal{U}(m,n)
\begin{bmatrix}
\psi_0(m,n)\\ \psi_1(m,n)
\end{bmatrix},
\qquad
\begin{bmatrix}
\psi_0(m,n+1)\\ \psi_1(m,n+1)
\end{bmatrix}
=\mathcal{V}(m,n)
\begin{bmatrix}
\psi_0(m,n)\\ \psi_1(m,n)
\end{bmatrix},
\end{gather}
where $\psi_i(m,n)=\psi(i,m,n)$ ($i=0,1$) and
the coef\/f\/icient matrices $\mathcal{U}(m,n)$, $\mathcal{V}(m,n)$ are
given by
\begin{gather*}
\mathcal{U}(m,n)=\begin{bmatrix}
\left\{1-a(0)/b\right\}u_0(m,n) & a(0)/b\\
\lambda a(1)/b & \left\{1-a(1)/b\right\}u_1(m,n)
\end{bmatrix},
\\
\mathcal{V}(m,n)=\begin{bmatrix}
\left\{1-a(0)/c\right\}v_0(m,n) & a(0)/c\\
\lambda a(1)/c & \left\{1-a(1)/c\right\}v_1(m,n)
\end{bmatrix},
\\
 u_\ell(m,n)=u(\ell,m,n),\qquad v_\ell(m,n)=v(\ell,m,n).
\end{gather*}
We remark that $u_\ell(m,n)$, $v_\ell(m,n)$ ($\ell=0,1$) obey the
constraints
\begin{gather}
u_0(m,n)u_1(m,n)=v_0(m,n)v_1(m,n)=1,
\label{2reduction_0}
\end{gather}
under the conditions
\eqref{2reduction_1}, \eqref{2reduction_2}.

The discrete Lax equation
\begin{gather}
\label{dLaxEq}
\mathcal{U}(m,n+1)\mathcal{V}(m,n)=\mathcal{V}(m+1,n)\mathcal{U}(m,n),
\end{gather}
follows from the compatibility condition for
\eqref{2x2LinearEqs_naHM_2-reduced}
and gives
\begin{gather}
 u_\ell(m,n)v_\ell(m+1,n)=u_\ell(m,n+1)v_\ell(m,n),
\label{2red_Eq1}\\
 \left\{a(\ell)-b\right\}u_\ell(m,n+1)-\left\{a(\ell)-c\right\}v_\ell(m+1,n)
\nonumber\\
\qquad {}= \left\{a(\ell+1)-b\right\}u_{\ell+1}(m,n)
-\left\{a(\ell+1)-c\right\}v_{\ell+1}(m,n),
\label{2red_Eq2}
\end{gather}
for $\ell=0,1$.
Using \eqref{2reduction_0} to eliminate the $u_1$, $v_1$ variables, one can solve \eqref{2red_Eq1}
and \eqref{2red_Eq2} for $u_0(m,n+1)$ and $v_0(m+1,n)$, to obtain the 2-reduced non-autonomous
discrete KP (na-dKP) equation,
\begin{gather}
u_0(m,n+1)=\frac{\tilde{F}(m,n)}{v_0(m,n)},\qquad
v_0(m+1,n)=\frac{\tilde{F}(m,n)}{u_0(m,n)},
\nonumber\\
\tilde{F}(m,n)=\frac{
\left\{a(1)-b\right\}v_0(m,n)
-\left\{a(1)-c\right\}u_0(m,n)
}{
\left\{a(0)-b\right\}u_0(m,n)
-\left\{a(0)-c\right\}v_0(m,n)
}.
\label{2reduced_dKP}
\end{gather}
Alternatively, we can use \eqref{2reduction_0}, \eqref{2red_Eq1} and \eqref{2red_Eq2}
to express $u_0(m,n+1)$, $v_0(m,n)$ in terms of $u_0(m,n)$, $v_0(m+1,n)$:
\begin{gather}
 u_0(m,n+1)=v_0(m+1,n)F(m,n), \qquad
v_0(m,n)=\frac{u_0(m,n)}{F(m,n)}, \nonumber\\
 F(m,n)=\frac{
\left\{a(1)-b\right\}+\left\{a(0)-c\right\}u_0(m,n)v_0(m+1,n)
}{
\left\{a(1)-c\right\}+\left\{a(0)-b\right\}u_0(m,n)v_0(m+1,n)
}.
\label{YBmap_AKP2}
\end{gather}
Motivated by \eqref{2reduced_dKP} and \eqref{YBmap_AKP2},
we def\/ine the following birational maps $\tilde{R}(b,c)$, $R(b,c)$:
\begin{gather}
\tilde{R}(b,c)  : \ (u,v)\mapsto \left(\frac{\tilde{F}}{v},\frac{\tilde{F}}{u}\right),\qquad
\tilde{F}=\frac{
\left\{a(1)-b\right\}v
-\left\{a(1)-c\right\}u
}{
\left\{a(0)-b\right\}u
-\left\{a(0)-c\right\}v
},\label{solitonic_map}\\
R(b,c)  : \  (u,v)\mapsto\left(v F,\frac{u}{F}\right),\qquad
F=\frac{
\left\{a(1)-b\right\}+\left\{a(0)-c\right\}uv
}{
\left\{a(1)-c\right\}+\left\{a(0)-b\right\}uv
}.\nonumber
\end{gather}
From now on we shall refer to the map $\tilde{R}(b,c)$ as the ``solitonic'' map, since
$\tilde{R}$ is nothing but the evolution rule for the discrete soliton equation~\eqref{2reduced_dKP}.
On the other hand, if we rescale as
$u\mapsto \left\{a(1)-b\right\}u$,
$v\mapsto \left\{a(1)-c\right\}v$,
the map $R(b,c)$ coincides with the map \eqref{dpKdV_map}
with $\kappa=\left\{a(0)-c\right\}\left\{a(1)-c\right\}$,
$\mu=\left\{a(0)-b\right\}\left\{a(1)-b\right\}$. Hence $R(b,c)$ is a genuine Yang--Baxter map, i.e.\ it satisf\/ies the Yang--Baxter
equation with parameters~\eqref{YBE}.
As discussed in \cite{SV}, this is a~direct consequence of the
3D-consistency of the discrete Lax equation~\eqref{dLaxEq}.
The solitonic map $\tilde{R}(b,c)$ satisf\/ies a similar but slightly dif\/ferent relation,
\begin{gather}
\tilde{R}_{12}(b,c)\tilde{R}_{13}(b,d)R_{23}(c,d)
=R_{23}(c,d)\tilde{R}_{13}(b,d)\tilde{R}_{12}(b,c),
\label{YBE_solitonic}
\end{gather}
which is also obtained from the 3D-consistency of the Lax pair.

\medskip

\noindent\textbf{Remark.}
Here we use the notation $\tilde{R}$
for ``solitonic'' maps and $R$ for Yang--Baxter maps,
whereas the opposite notation was used in \cite{KNW}.

\medskip

If we impose the $N$-reduction conditions,
\begin{gather}
u(\ell+N,m,n)=u(\ell,m,n),\qquad v(\ell+N,m,n)=v(\ell,m,n),
\nonumber\\
\psi(\ell+N,m,n)=\lambda\psi(\ell,m,n),\qquad
a(\ell+N)=a(\ell)\qquad \forall\, \ell\in\mathbb{Z},\label{N-reduction_1}\\
b(m)=b,\qquad c(n)=c \qquad \forall \, m,n\in\mathbb{Z},\nonumber
\end{gather}
instead of the 2-reduction conditions
\eqref{2reduction_1}, \eqref{2reduction_2},
the resulting Yang--Baxter map are related to
a geometric crystal of $A_{N-1}^{(1)}$-type
\cite{Etingof}.  Its ud-limit coincides with
a combinatorial $R$-matrix of $A_{N-1}^{(1)}$-type and hence with the
evolution rule for $A_{N-1}^{(1)}$-soliton cellular automata~\cite{HHIKTT}.

\section[From the Miwa equation to Yang-Baxter maps]{From the Miwa equation to Yang--Baxter maps}

In \cite{Miwa}, Miwa introduced
a discrete analogue of the bilinear BKP equation,
\begin{gather}
(a+b)(a+c)(b-c)\tau(100)\tau(011)
+(b+c)(b+a)(c-a)\tau(010)\tau(101)
\nonumber\\
\qquad {}+(c+a)(c+b)(a-b)\tau(001)\tau(110)
+(a-b)(b-c)(c-a)\tau(111)\tau(000)=0.
\label{MiwaEq}
\end{gather}
It is straightforward to generalize \eqref{MiwaEq} to the
non-autonomous case:
\begin{gather}
\left\{a(\ell)+b(m)\right\}\left\{a(\ell)+c(n)\right\}
\left\{b(m)-c(n)\right\}\tau(100)\tau(011)
\nonumber\\
\qquad{}
+\left\{b(m)+c(n)\right\}\left\{b(m)+a(\ell)\right\}
\left\{c(n)-a(\ell)\right\}\tau(010)\tau(101)
\nonumber\\
\qquad{}
+\left\{c(n)+a(\ell)\right\}\left\{c(n)+b(m)\right\}
\left\{a(\ell)-b(m)\right\}\tau(001)\tau(110)
\nonumber\\
\qquad{}
+\left\{a(\ell)-b(m)\right\}\left\{b(m)-c(n)\right\}
\left\{c(n)-a(\ell)\right\}\tau(111)\tau(000)=0.
\label{naMiwaEq}
\end{gather}
The non-autonomous Miwa (naM) equation \eqref{naMiwaEq}
arises as the compatibility condition for the linear system
(cf.~\cite{DJM_V,Nimmo_dBKP} for the autonomous case),
\begin{gather*}
\left\{a(\ell)+b(m)\right\}\psi(010)
-\left\{a(\ell)-b(m)\right\}u(000)\psi(110)
\nonumber\\
\qquad {}+\left\{a(\ell)-b(m)\right\}u(000)\psi(000)
-\left\{b(m)+a(\ell)\right\}\psi(100)=0,
\\
\left\{a(\ell)+c(n)\right\}\psi(001)
-\left\{a(\ell)-c(n)\right\}v(000)\psi(101)
\nonumber\\
\qquad {}+\left\{a(\ell)-c(n)\right\}u(000)\psi(000)
-\left\{c(n)+a(\ell)\right\}\psi(100)=0,
\end{gather*}
where $u(\ell,m,n)$, $v(\ell,m,n)$ are def\/ined as in~\eqref{def:uv}.

We now impose the $N$-reduction condition \eqref{N-reduction_1}
to obtain the discrete linear equations
\begin{gather}
\mathcal{U}_1(m,n)\begin{bmatrix}
\psi_0(m+1,n)\\ \vdots\\ \psi_{N-1}(m+1,n)
\end{bmatrix}
 =\mathcal{U}_2(m,n)
\begin{bmatrix}
\psi_0(m,n)\\ \vdots\\ \psi_{N-1}(m,n)
\end{bmatrix},
\label{BKP_linear_1}\\
\mathcal{V}_1(m,n)\begin{bmatrix}
\psi_0(m,n+1)\\ \vdots\\ \psi_{N-1}(m,n+1)
\end{bmatrix}
 =\mathcal{V}_2(m,n)
\begin{bmatrix}
\psi_0(m,n)\\ \vdots\\ \psi_{N-1}(m,n)
\end{bmatrix},
\label{BKP_linear_2}
\end{gather}
where $\psi_i(m,n)=\psi(i,m,n)$ ($i=0,1$) and
the coef\/f\/icient matrices $\mathcal{U}_i(m,n)$,
$\mathcal{V}_i(m,n)$ ($i=1,2$) are
given by
\begin{gather}
\mathcal{U}_1=\mathcal{I} + \mathcal{U}^{(0)}\Lambda,
\quad
\mathcal{U}_2=\mathcal{U}^{(0)}+\Lambda, \qquad
\mathcal{U}^{(0)}
=\mathrm{diag}\left[\beta_0u_0,\dots,\beta_{N-1}u_{N-1}\right],
\nonumber\\
\mathcal{V}_1=\mathcal{I} + \mathcal{V}^{(0)}\Lambda,
\qquad
\mathcal{V}_2=\mathcal{V}^{(0)}+\Lambda, \qquad
\mathcal{V}^{(0)}
=\mathrm{diag}\left[\gamma_0v_0,\dots,\gamma_{N-1}v_{N-1}\right],
\nonumber\\
u_\ell(m,n)=u(\ell,m,n),\qquad v_\ell(m,n)=v(\ell,m,n),
\nonumber\\
\Lambda=
\begin{bmatrix}
0 & 1 & & \\
 &0& \ddots & \\
 & &\ddots& 1\\
\lambda & & & 0
\end{bmatrix},
\qquad\beta_\ell=\frac{b-a(\ell)}{b+a(\ell)},
\qquad\gamma_\ell=\frac{c-a(\ell)}{c+a(\ell)},
\label{def:UV}
\end{gather}
where $\mathcal{I}$ is the identity matrix.
We remark that $u_\ell(m,n)$, $v_\ell(m,n)$ ($\ell=0,1,\dots,N-1$)
obey the constraints
\begin{gather}
u_0(m,n)\cdots u_{N-1}(m,n)=v_0(m,n)\cdots v_{N-1}(m,n)=1,
\label{N-reduction_0}
\end{gather}
under the conditions \eqref{N-reduction_1}.
The compatibility condition of
\eqref{BKP_linear_1} and \eqref{BKP_linear_2} is
\begin{gather}
\mathcal{U}_1(m,n+1)^{-1}\mathcal{U}_2(m,n+1)
\mathcal{V}_1(m,n)^{-1}\mathcal{V}_2(m,n)\nonumber\\
\qquad {}=\mathcal{V}_1(m+1,n)^{-1}\mathcal{V}_2(m+1,n)
\mathcal{U}_1(m,n)^{-1}\mathcal{U}_2(m,n).\label{dLax_BKP_N-reduced}
\end{gather}
We remark that $\det\mathcal{U}_1 = \det\mathcal{V}_1
 = 1+(-1)^{N-1}\lambda\prod_{j=0}^{N-1}\beta_j$ and
hence $\mathcal{U}_1$, $\mathcal{V}_1$ are invertible.

To write down the Yang--Baxter map and the solitonic map
associated with the discrete Lax equation~\eqref{dLax_BKP_N-reduced},
we use the abbreviations,
$u_j:=u_j(0,0)$, $v_j:=v_j(0,0)$,
$\bar{u}_j:=u_j(0,1)$, $\bar{v}_j:=v_j(1,0)$,
$\bm{u}=(u_0,\dots,u_{N-1})$, $\bm{v}=(v_0,\dots,v_{N-1})$,
$\bar{\bm{u}}=(\bar{u}_0,\dots,\bar{u}_{N-1})$,
$\bar{\bm{v}}=(\bar{v}_0,\dots,\bar{v}_{N-1})$.
Using these abbreviations, we have the following
discrete evolution equations:
\begin{gather}
\bar{u}_j=u_j
\frac{f_{j+1}(\bm{u},\bm{v})}{f_j(\bm{u},\bm{v})},\qquad
\bar{v}_j=v_j\frac{f_{j+1}(\bm{u},\bm{v})}{f_j(\bm{u},\bm{v})},\qquad
j=0,1,\dots,N-1,\nonumber
\\
f_j(\bm{u},\bm{v})
=\sum_{k=0}^{N-1}
\prod_{l=0}^{k-1}\beta_{j+l}\gamma_{j+l}u_{j+l}v_{j+l}
\left(\beta_{j+k}u_{j+k}-\gamma_{j+k}v_{j+k}\right).
\label{dBKPeq_N-reduced}
\end{gather}
Here and in what follows, the subscripts are considered as elements of
$\mathbb{Z}/N\mathbb{Z}$.

One can show that
the discrete Lax equation \eqref{dLax_BKP_N-reduced} is
equivalent to
\begin{gather}
\bar{u}_j
=u_j\frac{g_j(\bm{u},\bar{\bm{v}})}{g_{j+1}(\bm{u},\bar{\bm{v}})},
\qquad v_j
=\bar{v}_j\frac{g_{j+1}(\bm{u},\bar{\bm{v}})}{g_j(\bm{u},\bar{\bm{v}})},
\qquad j=0,1,\dots,N-1,
\nonumber\\
 g_j(\bm{u},\bar{\bm{v}})=
\sum_{k=0}^{N-1}
\left\{\prod_{l=0}^{k-1}\gamma_{j+l}\bar{v}_{j+l}
\cdot (\beta_{j+k}\gamma_{j+k}u_{j+k}\bar{v}_{j+k}-1)\cdot
\prod_{l=k+1}^{N-1}\beta_{j+l}u_{j+l}\right\}.
\label{def:YBM_BKP}
\end{gather}
We will give a derivation of the equations
\eqref{dBKPeq_N-reduced}, \eqref{def:YBM_BKP} in
Appendix~\ref{appendix:proof}.

In analogy with \eqref{solitonic_map} we shall refer to the equations \eqref{dBKPeq_N-reduced}
as a solitonic map:
\begin{gather}
\begin{array}{@{}ccccc}
\tilde{R}(\bm{\beta},\bm{\gamma}) &:& \mathbb{C}^N\times\mathbb{C}^N &
\to & \mathbb{C}^N\times\mathbb{C}^N\\
&& \rotatebox{90}{$\in$} && \rotatebox{90}{$\in$}\\
&& (\bm{u},\bm{v}) & \mapsto & (\bar{\bm{u}},\bar{\bm{v}})
\end{array}.
\label{solitonicMap_BKP}
\end{gather}
The corresponding Yang--Baxter map is obtained from \eqref{def:YBM_BKP}:
\begin{gather}
\begin{array}{@{}ccccc}
R(\bm{\beta},\bm{\gamma}) &:& \mathbb{C}^N\times\mathbb{C}^N &
\to & \mathbb{C}^N\times\mathbb{C}^N\\
&& \rotatebox{90}{$\in$} && \rotatebox{90}{$\in$}\\
&& (\bm{u},\bar{\bm{v}}) & \mapsto & (\bar{\bm{u}},\bm{v})
\end{array}.
\label{YBM_BKP}
\end{gather}
The birational map \eqref{YBM_BKP} satisf\/ies the
Yang--Baxter equation \eqref{YBE}, whereas when considered together, the Yang--Baxter map \eqref{YBM_BKP} and
the BKP solitonic map \eqref{solitonicMap_BKP}
satisfy \eqref{YBE_solitonic}.
We remark again that these ``Yang--Baxter'' properties are
mere consequences of the 3D consistency of
the discrete Lax equation \eqref{dLax_BKP_N-reduced}.

Next we consider the ud-limit of the Yang--Baxter map
\eqref{def:YBM_BKP}, \eqref{YBM_BKP}.
To do this, we replace the parameters $\beta$ and $\gamma$ by
\begin{gather*}
\beta_j\mapsto\sqrt{-1}\beta_j, \qquad
\gamma_j\mapsto\sqrt{-1}\gamma_j, \qquad j=0,1,\dots,N-1,
\end{gather*}
and introduce new independent variables $U_j$, $V_j$, $\bar{U}_j$, $\bar{V}_j$ as def\/ined in
\begin{gather*}
e^{U_j/\epsilon}=\beta_ju_j,\qquad
e^{V_j/\epsilon}=\gamma_jv_j \qquad
e^{\bar{U}_j/\epsilon}=\beta_j\bar{u}_j,\qquad
e^{\bar{V}_j/\epsilon}=\gamma_j\bar{v}_j, \qquad j=0,1,\dots,N-1.
\end{gather*}
It follows from \eqref{N-reduction_0} that
$\sum_{j=0}^{N-1}U_j=\sum_{j=0}^{N-1}\bar{U}_j$ and
$\sum_{j=0}^{N-1}V_j=\sum_{j=0}^{N-1}\bar{V}_j$ are
independent of $m$ and $n$.
Then taking the ud-limit of \eqref{def:YBM_BKP} gives
\begin{gather}
\bar{U}_j
=U_j+G_j(\bm{U},\bar{\bm{V}})-G_{j+1}(\bm{U},\bar{\bm{V}}),
\nonumber\\
V_j
=\bar{V}_j+G_{j+1}(\bm{U},\bar{\bm{V}})-G_j(\bm{U},\bar{\bm{V}}),
\qquad j=0,1,\dots,N-1,
\nonumber\\
G_j(\bm{U},\bar{\bm{V}})=
\max_{0\le k<N}
\left(\sum_{l=0}^{k-1}\bar{V}_{j+1}
+ \max(0,U_{j+k}+\bar{V}_{j+k})+
\sum_{l=k+1}^{N-1}U_{j+l}\right),\label{def:udYBM_BKP}
\\
\bm{U}=(U_0,\dots,U_{N-1}),\qquad
\bm{V}=(V_0,\dots,V_{N-1}),
\nonumber\\
\bar{\bm{U}}=(\bar{U}_0,\dots,\bar{U}_{N-1}),\qquad
\bar{\bm{V}}=(\bar{V}_0,\dots,\bar{V}_{N-1}),\nonumber
\end{gather}
which def\/ines a piecewise-linear Yang--Baxter map $R^\mathrm{(ud)}$:
\begin{gather*}
\begin{array}{@{}ccccc}
R^\mathrm{(ud)} &:& \mathbb{R}^N &\to& \mathbb{R}^N\\
&& \rotatebox{90}{$\in$} && \rotatebox{90}{$\in$}\\
&& \left(\bm{U},\bar{\bm{V}}\right)
&\mapsto&
\left(\bar{\bm{U}},\bm{V}\right)
\end{array}.
\end{gather*}
We remark that if we impose the condition $U_j, \bar{V}_j\ge 0$
($j=0,1,\dots,N-1$), the Yang--Baxter map $R^{\mathrm{(ud)}}$
coincides with the combinatorial $R$-matrix of type $A_{N-1}^{(1)}$
(see~(2.12) in~\cite{Takagi}).

\begin{figure}[t]
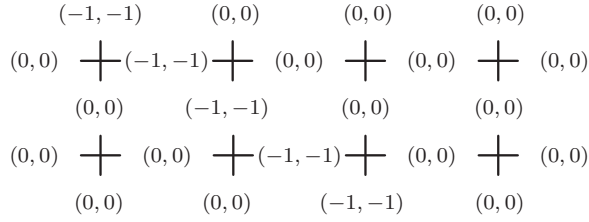

\begin{center}
\hspace{-2mm}{\uvec{-1,-1}}\hspace{9mm}\uvec{0,0}%
\hspace{11mm}\uvec{0,0}\hspace{11mm}\uvec{0,0}\\[1.5mm]
\uvec{0,0}\hspace{2.35mm}\lplus\uvec{-1,-1}\lplus\emp%
\lplus\emp\lplus\hspace{2.35mm}\uvec{0,0}\\[1.5mm]
\uvec{0,0}\hspace{8mm}{\uvec{-1,-1}}%
\hspace{9.5mm}\uvec{0,0}\hspace{11.1mm}\uvec{0,0}\\[1.5mm]
\uvec{0,0}\hspace{2.35mm\lplus\emp}\lplus\uvec{-1,-1}%
\lplus\emp\lplus\emp\\[1.5mm]
\uvec{0,0}\hspace{10.35mm}\uvec{0,0}%
\hspace{9mm}{\uvec{-1,-1}}\hspace{9.35mm}\uvec{0,0}
\caption{1-soliton propagation.}
\label{fig:1-soliton}
\end{center}
\end{figure}

\begin{figure}[t]
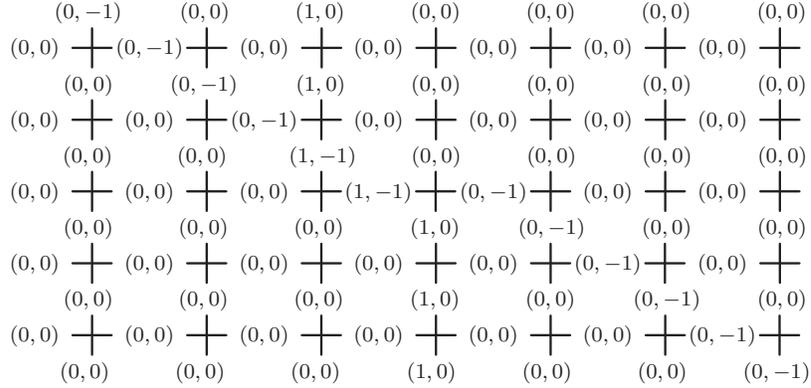

\begin{center}
\hspace{7.5mm}{\uvec{0,-1}}\hspace{7.7mm}\uvec{0,0}%
\hspace{8.7mm}{\uvec{1,0}}\hspace{8.7mm}\uvec{0,0}%
\hspace{8.7mm}\uvec{0,0}\hspace{8.7mm}\uvec{0,0}%
\hspace{8.7mm}\uvec{0,0}\\
\empp\lplus\uvec{0,-1}\lplus\empp%
\lplus\empp\lplus\empp\lplus\empp\lplus\empp\lplus\\
\hspace{8.7mm}\uvec{0,0}\hspace{7.6mm}{\uvec{0,-1}}%
\hspace{7.7mm}{\uvec{1,0}}\hspace{8.7mm}\uvec{0,0}%
\hspace{8.7mm}\uvec{0,0}\hspace{8.7mm}\uvec{0,0}%
\hspace{8.7mm}\uvec{0,0}\\
\empp\lplus\empp\lplus\uvec{0,-1}\lplus\empp%
\lplus\empp\lplus\empp\lplus\empp\lplus\\
\hspace{8.7mm}\uvec{0,0}\hspace{8.6mm}\uvec{0,0}%
\hspace{8.2mm}{\uvec{1,-1}}\hspace{7.3mm}\uvec{0,0}%
\hspace{8.7mm}\uvec{0,0}\hspace{8.7mm}\uvec{0,0}\hspace{8.7mm}\uvec{0,0}\\
\empp\lplus\empp\lplus\empp\lplus\uvec{1,-1}%
\lplus\uvec{0,-1}\lplus\empp\lplus\empp\lplus\\
\hspace{8.7mm}\uvec{0,0}\hspace{8.7mm}\uvec{0,0}%
\hspace{8.7mm}\uvec{0,0}\hspace{8.8mm}{\uvec{1,0}}%
\hspace{7.7mm}{\uvec{0,-1}}%
\hspace{7.5mm}\uvec{0,0}\hspace{8.7mm}\uvec{0,0}\\
\empp\lplus\empp\lplus\empp\lplus\empp%
\lplus\empp\lplus\uvec{0,-1}\lplus\empp\lplus\\
\hspace{8.7mm}\uvec{0,0}\hspace{8.7mm}\uvec{0,0}%
\hspace{8.7mm}\uvec{0,0}\hspace{8.8mm}{\uvec{1,0}}%
\hspace{8.7mm}\uvec{0,0}\hspace{7.7mm}{\uvec{0,-1}}%
\hspace{7.5mm}\uvec{0,0}\\
\empp\lplus\empp\lplus\empp\lplus\empp%
\lplus\empp\lplus\empp\lplus\uvec{0,-1}\lplus\\
\hspace{8.7mm}\uvec{0,0}\hspace{8.7mm}\uvec{0,0}%
\hspace{8.7mm}\uvec{0,0}\hspace{8.8mm}{\uvec{1,0}}%
\hspace{8.7mm}\uvec{0,0}\hspace{8.7mm}\uvec{0,0}%
\hspace{7.4mm}{\uvec{0,-1}}
\caption{2-soliton interaction.}
\label{fig:2-soliton}
\end{center}
\end{figure}

However, the piecewise-linear map \eqref{def:udYBM_BKP} also admits
soliton-type solutions in the general case. To describe these solitons, we use
the notation of Fig.~\ref{fig:ud-R}.
Figs.~\ref{fig:1-soliton} and \ref{fig:2-soliton} show
soliton-type phenomena in the 3-reduced case, where we have
set $U_0+U_1+U_2=\bar{U}_0+\bar{U}_1+\bar{U}_2=1$ and
$\bar{V}_0+\bar{V}_1+\bar{V}_2=V_0+V_1+V_2=0$. The evolution shown in Fig.~\ref{fig:2-soliton}
is reminiscent of the interactions of solitons with stationary solutions that have been studied by Hirota in case of
the ultra-discrete limit of the discrete Sawada--Kotera equation \cite{Hirota_ud-dSK}.

\begin{figure}[t]
\begin{center}
$R$:
\raisebox{3mm}{\begin{tabular}{c}
{\small $(U_0,U_1)$}\\[1mm]
\raisebox{1.5mm}{\small $(\bar{V}_0,\bar{V}_1)$}%
\ {\Huge $+$}%
\ \raisebox{1.5mm}{\small $(V_0,V_1)$}\\
{\small $(\bar{U}_0,\bar{U}_1)$}\\[-1.25cm]
\rotatebox{-45}{\Large $\Rightarrow$}
\end{tabular}}
\ \\[4.25mm]
\caption{Graphical representation.}
\label{fig:ud-R}
\end{center}
\end{figure}

\section{Concluding remarks}

In this article, we have constructed
rational and piecewise-linear Yang--Baxter maps from
the $N$-reduction of the discrete BKP equation.
It should be remarked that the 4D consistency property~\cite{MN}
of the (non-reduced) BKP equation was observed in~\cite{ABS}.

{}From the Lie-algebraic viewpoint,
the $(2n+1)$-reduced BKP equation has the symmetry of
the af\/f\/ine Lie algebra of $A_{2n}^{(2)}$-type and
the $(2n)$-reduced one has that of $D_{n+1}^{(2)}$~\cite{DJKM}.
It is to be expected that the reduced ultra-discrete BKP system~\eqref{def:udYBM_BKP} is somehow related to
crystal bases of  the above types.
However, in the 3-reduced case,
our piecewise-linear map~\eqref{def:udYBM_BKP} seems
to be dif\/ferent from the combinatorial $R$-matrix of
$A_{2}^{(2)}$-type, discussed explicitly in~\cite{HKT}, as~\eqref{def:udYBM_BKP} has a rotational symmetry which
the $A_{2}^{(2)}$-type map in~\cite{HKT} does not possess.
The exact relationship between the $N$-reduced discrete BKP equation
and $A_{2}^{(2)}$-type crystal bases needs further clarif\/ication.

\appendix

\section[Derivation of the $N$-reduced BKP equation and
related Yang-Baxter map]{Derivation of the $\boldsymbol{N}$-reduced BKP equation\\ and
related Yang--Baxter map}
\label{appendix:proof}

Here we give a derivation of the $N$-reduced BKP equation
\eqref{dBKPeq_N-reduced} and the related Yang--Baxter map \eqref{def:YBM_BKP}.

In the case $N=2$, the nonlinear equations associated
with the discrete Lax equation \eqref{dLax_BKP_N-reduced}
are the same as in the 2-reduced $A$-case, i.e.,
those in Section \ref{section:dAKP} (as is to be expected for Lie-algebraic reasons).
The f\/irst non-trivial example is the 3-reduced case and
hereafter we assume $N\ge 3$.

If $\beta_j\ne 0$, $\gamma_j\ne 0$ ($j=0,1,\dots,N-1$) in \eqref{def:UV},
we can set $\beta_j=\gamma_j=1$ ($j=0,1,\dots,N-1$)
by rescaling the variables $\{u_j\}$, $\{v_j\}$.
In this appendix, we choose $\beta_j=\gamma_j=1$.

We f\/irst remark that the inverse of $\mathcal{U}_1$ of \eqref{def:UV}
is of the form,
\begin{gather*}
\mathcal{U}_1^{-1}
=\left\{1+(-1)^{N-1}\lambda\prod_{i=0}^{N-1}\beta_i\right\}^{-1}
\left\{
\mathcal{I}+\sum_{j=1}^{N-1}(-1)^j
\left(\prod_{k=0}^{j-1}\mathcal{U}^{(k)}\right)\Lambda^j
\right\}
\end{gather*}
where the matrices
\begin{gather*}
\mathcal{U}^{(k)} = \mathrm{diag}\left[u_{k},u_{k+1},\dots,u_{k+N-1}\right],
\qquad k=0,1,\dots,N-1,
\end{gather*}
have the properties,
\begin{gather}
\mathcal{U}^{(k+N)}=\mathcal{U}^{(k)},\qquad
\Lambda\mathcal{U}^{(k)}=\mathcal{U}^{(k+1)}\Lambda,
\qquad k=0,1,\dots,N-1.
\label{invU1}
\end{gather}
We then introduce a degree on
$\mathrm{Mat}(N)$, the space of $N\times N$ matrices:
\begin{gather}
\forall\, X\in\mathrm{Mat}(N)\qquad
\mathrm{deg}(X)=n
\ \Leftrightarrow\
\left[d,X\right]=nX, \qquad
d=N\lambda\frac{d}{d\lambda}-\mathrm{diag}\left[1,2,\dots,N\right].\!\!
\label{def:degree}
\end{gather}
Substituting \eqref{invU1} into \eqref{dLax_BKP_N-reduced} and
arranging in order of the degree \eqref{def:degree}, we obtain
a set of equations for $u_j$, $v_j$, $\bar{u}_j$, $\bar{v}_j$.
To write down these equations, we prepare some additional notation:
\begin{gather*}
\bm{\mathcal{U}} = \big(
\mathcal{U}^{(0)},\mathcal{U}^{(1)},\dots,\mathcal{U}^{(N-1)}
\big),
\qquad
\bm{\mathcal{V}} = \big(
\mathcal{V}^{(0)},\mathcal{V}^{(1)},\dots,\mathcal{V}^{(N-1)}
\big),
\\
\bar{\bm{\mathcal{U}}} = \big(
\bar{\mathcal{U}}^{(0)},\bar{\mathcal{U}}^{(1)},
\dots,\bar{\mathcal{U}}^{(N-1)}
\big),
\qquad
\bar{\bm{\mathcal{V}}} = \big(
\bar{\mathcal{V}}^{(0)},\bar{\mathcal{V}}^{(1)},
\dots,\bar{\mathcal{V}}^{(N-1)}
\big),
\end{gather*}
where $\bar{\mathcal{U}}^{(j)}=\mathcal{U}^{(j)}(0,1)$,
$\bar{\mathcal{V}}^{(j)}=\mathcal{V}^{(j)}(1,0)$.
We denote by
$h_k\left(\bar{\bm{\mathcal{U}}},\bm{\mathcal{V}}\right)\Lambda^k$
the degree $k$ terms in
$(\det\bar{\mathcal{U}}_1)(\det\mathcal{V}_1)
\bar{\mathcal{U}}_1^{-1}\bar{\mathcal{U}}_2
\mathcal{V}_1^{-1}\mathcal{V}_2$,
i.e.,
\begin{gather*}
(\det\bar{\mathcal{U}}_1)(\det\mathcal{V}_1)
\bar{\mathcal{U}}_1^{-1}\bar{\mathcal{U}}_2
\mathcal{V}_1^{-1}\mathcal{V}_2
=\sum_{k=0}^{2N}
h_k\left(\bar{\bm{\mathcal{U}}},\bm{\mathcal{V}}\right)\Lambda^k.
\end{gather*}
For example,
$h_0\left(\bar{\bm{\mathcal{U}}},\bm{\mathcal{V}}\right)$,
$h_1\left(\bar{\bm{\mathcal{U}}},\bm{\mathcal{V}}\right)$
are given by
\begin{gather*}
h_0\left(\bar{\bm{\mathcal{U}}},\bm{\mathcal{V}}\right)
= \bar{\mathcal{U}}^{(0)}\mathcal{V}^{(0)},
\\
h_1\left(\bar{\bm{\mathcal{U}}},\bm{\mathcal{V}}\right)
=
\bar{\mathcal{U}}^{(0)}
\big(\mathcal{I}-\mathcal{V}^{(0)}\mathcal{V}^{(1)}\big)
+\big(\mathcal{I}-\bar{\mathcal{U}}^{(0)}\bar{\mathcal{U}}^{(1)}\big)
\mathcal{V}^{(1)}.
\end{gather*}
By using this notation, the discrete Lax equation
\eqref{dLax_BKP_N-reduced} can be rewritten as
\begin{gather}
h_k\left(\bar{\bm{\mathcal{U}}},\bm{\mathcal{V}}\right)
=h_k\left(\bar{\bm{\mathcal{V}}},\bm{\mathcal{U}}\right),
\qquad k=0,1,\dots,2N.
\label{h=h}
\end{gather}

The system of polynomial equations \eqref{h=h} is of course overdetermined.
However, it has a~unique nontrivial solution.
For the moment, we consider only two equations,
\begin{gather}
h_0\left(\bar{\bm{\mathcal{U}}},\bm{\mathcal{V}}\right)
=h_0\left(\bar{\bm{\mathcal{V}}},\bm{\mathcal{U}}\right),
\qquad
h_1\left(\bar{\bm{\mathcal{U}}},\bm{\mathcal{V}}\right)
=h_1\left(\bar{\bm{\mathcal{V}}},\bm{\mathcal{U}}\right).
\label{eqs_deg0_deg1}
\end{gather}
To solve \eqref{eqs_deg0_deg1}, we introduce another set of
variables \cite{Etingof,SV}:
\begin{gather*}
x_j= \prod_{i=0}^j v_k, \qquad \bar{x}_j= \prod_{i=0}^j \bar{v}_k,
\qquad j=0,1,\dots,N-2.
\end{gather*}
It is straightforward to show that
$\vec{x}
={}^t\left[1:x_0:\cdots:x_{N-2}\right]\in\mathbb{P}^{N-1}$
satisf\/ies a linear equation,
\begin{gather}
M\vec{x}=\vec{0},
\label{Mx=0}\\
M=\begin{bmatrix}
-u_0\bar{v}_0p_1 & q_0 & -p_0 & & & \\
 & -u_1\bar{v}_1p_2 & q_1 & -p_1 & & \\
 & & \ddots & \ddots & \ddots & \\
& & & -u_{N-3}\bar{v}_{N-3}p_{N-2} & q_{N-3} & -p_{N-3}\\
-p_{N-2}V & & & & -u_{N-2}\bar{v}_{N-2}p_{N-1} & q_{N-2}\\
q_{N-1}V&-p_{N-1}V & & & & -u_{N-1}\bar{v}_{N-1}p_0
\end{bmatrix}
\nonumber
\end{gather}
where $p_j=u_j\bar{v}_j-1$,
$q_j=p_ju_{j+1}+\bar{v}_jp_{j+1}$ ($j=0,1,\dots,N-1$),
and $V=\prod_{k=0}^{N-1}v_k=\prod_{k=0}^{N-1}\bar{v}_k$.

\begin{theorem}
\label{theorem:solution}
Assume that
\begin{gather}
\label{generic_case}
u_j\bar{v}_j\ne 1, \qquad
\sum_{k=0}^n
\left\{\prod_{l=0}^{k-1}\bar{v}_{j+l}
\cdot (u_{j+k}\bar{v}_{j+k}-1)\cdot\!\prod_{l=k+1}^{n}\!u_{j+l}\right\}
\ne 0,\!
\qquad j=0,1,\dots,N-1.\!\!\!
\end{gather}
Under this assumption,
the space of solutions for the linear equation~\eqref{Mx=0}
is one-dimensional, and a basis is given by
\begin{gather}
{}^t\left[1:x_0:x_1:\cdots:x_{N-2}\right]
={}^t\left[g_0:\bar{v}_0g_1:\bar{v}_0\bar{v}_1g_2:\cdots:
(\bar{v}_0\cdots\bar{v}_{N-2})g_{N-1}\right],
\label{solution_x}
\end{gather}
where $g_j=g_j(\bm{u},\bar{\bm{v}})$
$(j=0,1,\dots,N-1)$ are defined as \eqref{def:YBM_BKP}.
\end{theorem}
To prove this theorem, we prepare two lemmas.

\begin{lemma}
\label{lemma:Dn}
Let $D_n(m)$ be the determinant of order $n$:
\begin{gather*}
D_n(m)=
\begin{vmatrix}
q_m & -p_m & &\\
-u_{m+1}\bar{v}_{m+1}p_{m+2} & q_{m+1} & \ddots &\\
& \ddots & \ddots & -p_{m+n-2}\\
     & & -u_{m+n-1}\bar{v}_{m+n-1}p_{m+n} & q_{m+n-1}
\end{vmatrix}.
\end{gather*}
Then $D_n(m)$ can be expressed as
\begin{gather}
\label{Dn(m)}
D_n(m)=\prod_{i=1}^{n-1}(u_{m+i}\bar{v}_{m+i}-1)\cdot
\sum_{j=0}^n
\left\{\prod_{k=0}^{j-1}\bar{v}_{m+k}
\cdot (u_{m+j}\bar{v}_{m+j}-1)\cdot
\prod_{k=j+1}^{n}u_{m+k}\right\}.
\end{gather}
\end{lemma}
\begin{proof}
It is enough to prove the case $m=0$.
Using the recurrence relation
$D_{n+2}(0)=q_{n+1}D_{n+1}(0)-u_{n+1}\bar{v}_{n+1}p_{n}p_{n+2}D_{n}(0)$,
one can prove the Lemma~\ref{lemma:Dn} by induction on~$n$.
\end{proof}

\begin{lemma}
\label{lemma:detM=0}
$\det M=0$.
\end{lemma}
\begin{proof}
The desired result follows from \eqref{Dn(m)} and a judicious expansion of $M$ involving the $(N-1)$-th and $N$-th rows.
\end{proof}

\begin{proof}[Proof of Theorem \ref{theorem:solution}]
Denote by $M_{N}^1$ the matrix
that results from $M$ by deleting the $N$-th row and
the f\/irst column.
Under the assumption \eqref{generic_case},
$
\det M_{N}^1=D_{N-1}(0)
\ne 0
$.
Together with $\det M=0$ (Lemma \ref{lemma:detM=0}),
it follows that $\dim\mathrm{Ker}\,M=1$.
The solution \eqref{solution_x} can be checked by direct
substitution into \eqref{Mx=0}.
\end{proof}

Thus we have obtained \eqref{def:YBM_BKP}.
Next we will show that every relation in \eqref{h=h} holds
if the variables $\bm{u}$, $\bm{v}$, $\bar{\bm{u}}$, $\bar{\bm{v}}$
satisfy the relations~\eqref{def:YBM_BKP}.

For $\bm{a}=(a_0,\dots,a_{N-1})$, $\bm{b}=(b_0,\dots,b_{N-1})$,
we def\/ine polynomials
$g_j^{(m)}(\bm{a},\bm{b})$ ($j,m=0,1,\dots$, $N-1$) as
\begin{gather*}
g^{(m)}_j(\bm{a},\bm{b})
=\sum_{k=0}^{m}
\left\{\prod_{l=0}^{k-1}a_{j+l}
\cdot (a_{j+k}b_{j+k}-1)\cdot\prod_{l=k+1}^{m}b_{j+l}\right\}.
\end{gather*}

\begin{lemma}
\label{lemma:covariant_quantities}
If the variables $\bm{u}$, $\bm{v}$, $\bar{\bm{u}}$, $\bar{\bm{v}}$
obey the relation~\eqref{def:YBM_BKP}, then the following relations hold:
\begin{gather*}
g^{(m)}_j\left(\bar{\bm{u}},\bm{v}\right)
=g^{(m)}_j\left(\bar{\bm{v}},\bm{u}\right),
\qquad j,m=0,1,\dots,N-1.
\end{gather*}
\end{lemma}
\begin{proof}
Since the equation is invariant under the rotation
of the suf\/f\/ices, i.e.\
$0\mapsto 1\mapsto 2\mapsto \cdots\mapsto N-1\mapsto 0$,
one can set $j=0$ without loss of generality.
The desired result can be proved
by induction on $m$, using
the recurrence relation
\begin{gather*}
g^{(m+1)}_0\left(\bm{a},\bm{b}\right)
=g^{(m)}_0\left(\bm{a},\bm{b}\right)b_m
+(a_0\cdots a_{m-1})(a_mb_m-1).\tag*{\qed}
\end{gather*}
\renewcommand{\qed}{}
\end{proof}

A straightforward calculation shows that
\begin{gather}
h_0\left(\bar{\bm{\mathcal{U}}},\bm{\mathcal{V}}\right)
=g_0^{(0)}\left(\bar{\bm{\mathcal{U}}},\bm{\mathcal{V}}\right)
+\mathcal{I},
\nonumber\\
h_1\left(\bar{\bm{\mathcal{U}}},\bm{\mathcal{V}}\right)
=-g_0^{(1)}\left(\bar{\bm{\mathcal{U}}},\bm{\mathcal{V}}\right),
\nonumber\\
h_j\left(\bar{\bm{\mathcal{U}}},\bm{\mathcal{V}}\right)
=(-1)^j
\left\{
g_0^{(j)}\left(\bar{\bm{\mathcal{U}}},\bm{\mathcal{V}}\right)
+g_0^{(j-2)}\left(\bar{\bm{\mathcal{U}}},\bm{\mathcal{V}}\right)
\right\},
\qquad 2\le j\le N-1,
\nonumber\\
h_N\left(\bar{\bm{\mathcal{U}}},\bm{\mathcal{V}}\right)
=(-1)^N\left\{
\bar{\mathcal{U}}^{(0)}\mathcal{V}^{(0)}
g_1^{(N-2)}\left(\bar{\bm{\mathcal{U}}},\bm{\mathcal{V}}\right)
-g_0^{(N-2)}\left(\bar{\bm{\mathcal{U}}},\bm{\mathcal{V}}\right)
-\prod_{i=0}^{N-1}\bar{\mathcal{U}}^{(i)}
-\prod_{i=0}^{N-1}\mathcal{V}^{(i)}
\right\},
\nonumber\\
h_{2N-k}\left(\bar{\bm{\mathcal{U}}},\bm{\mathcal{V}}\right)
=(-1)^k
\prod_{i=0}^{N-k}\big(
\bar{\mathcal{U}}^{(i)}\mathcal{V}^{(i)}
\big)
g_{N-k+1}^{(k-2)}\left(\bar{\bm{\mathcal{U}}},\bm{\mathcal{V}}\right)\label{h_to_g}
\\
\phantom{h_{2N-k}\left(\bar{\bm{\mathcal{U}}},\bm{\mathcal{V}}\right)}
{}+(-1)^{k+1}\prod_{i=0}^{N-k-2}\big(
\bar{\mathcal{U}}^{(i)}\mathcal{V}^{(i)}
\big)
g_{N-k-1}^{(k)}\left(\bar{\bm{\mathcal{U}}},\bm{\mathcal{V}}\right)
\quad (2\le k\le N-1),
\nonumber\\
h_{2N-1}\left(\bar{\bm{\mathcal{U}}},\bm{\mathcal{V}}\right)
=
\prod_{i=0}^{N-3}\big(
\bar{\mathcal{U}}^{(i)}\mathcal{V}^{(i)}
\big)
g_{N-2}^{(1)}\left(\bar{\bm{\mathcal{U}}},\bm{\mathcal{V}}\right),
\nonumber\\
h_{2N}\left(\bar{\bm{\mathcal{U}}},\bm{\mathcal{V}}\right)
=
\prod_{i=0}^{N-2}\big(
\bar{\mathcal{U}}^{(i)}\mathcal{V}^{(i)}
\big).
\nonumber
\end{gather}

Lemma \ref{lemma:covariant_quantities} and relation~\eqref{h_to_g}
imply that the overdetermined system \eqref{h=h} is satisf\/ied
if the variables $\bm{u}$, $\bm{v}$, $\bar{\bm{u}}$, $\bar{\bm{v}}$
obey the relation~\eqref{def:YBM_BKP}.
In other words, this means that the discrete Lax equation~\eqref{dLax_BKP_N-reduced} is equivalent to~\eqref{def:YBM_BKP}.

Just as the Yang--Baxter map \eqref{def:YBM_BKP} was shown to be equivalent
to~\eqref{eqs_deg0_deg1}, under the condition~\eqref{generic_case},
the solitonic map~\eqref{dBKPeq_N-reduced} can also be shown to be equivalent to~\eqref{eqs_deg0_deg1}.
We omit however the details of this proof.

\subsection*{Acknowledgments}
The authors are grateful to Professors Atsuo Kuniba, Masato Okado, and
Yasuhiko Yamada for discussions and comments.
S.K.\ wishes to acknowledge support from the Japan Society for the
Promotion of Science (JSPS) through a Grant-In-Aid for Scientif\/ic
Research (No.~19540228);
R.W.\ also acknowledges support by JSPS through a Grant-in-Aid (No.~21540210) and J.J.C.N.\ and R.W.\ acknowledge f\/inancial support from the British Council (PMI2 Research Co-operation award).

\pdfbookmark[1]{References}{ref}
\LastPageEnding

\end{document}